\documentclass[
reprint,
secnumarabic,
amssymb, amsmath,
aps,prl,
groupedaddress,
frontmatterverbose,
]{revtex4-1}
\usepackage{amsmath}
\usepackage{graphicx}
\usepackage{docs}%
\usepackage{bm}%
\usepackage{mathtools}
\usepackage[colorlinks=true,linkcolor=blue]{hyperref}%
\expandafter\ifx\csname package@font\endcsname\relax\else
\expandafter\expandafter
\expandafter\usepackage
\expandafter\expandafter
\expandafter{\csname package@font\endcsname}%
\fi
\begin{document}

\title{Short-Term Plasticity and Long-Term Potentiation in Magnetic Tunnel Junctions: Towards Volatile Synapses}%

\author{Abhronil Sengupta}
\email{asengup@purdue.edu}
\author{Kaushik Roy}

\affiliation{School of Electrical \& Computer Engineering, Purdue University, West Lafayette, IN 47907, USA}%

\date{October 2015}%

\begin{abstract}
Synaptic memory is considered to be the main element responsible for learning and cognition in humans. Although traditionally non-volatile long-term plasticity changes have been implemented in nanoelectronic synapses for neuromorphic applications, recent studies in neuroscience have revealed that biological synapses undergo meta-stable volatile strengthening followed by a long-term strengthening provided that the frequency of the input stimulus is sufficiently high. Such ``memory strengthening'' and ``memory decay'' functionalities can potentially lead to adaptive neuromorphic architectures. In this paper, we demonstrate the close resemblance of the magnetization dynamics of a Magnetic Tunnel Junction (MTJ) to short-term plasticity and long-term potentiation observed in biological synapses. We illustrate that, in addition to the magnitude and duration of the input stimulus, frequency of the stimulus plays a critical role in determining long-term potentiation of the MTJ. Such MTJ synaptic memory arrays can be utilized to create compact, ultra-fast and low power intelligent neural systems. 
\end{abstract}

\maketitle
\section{Introduction}
With significant research efforts being directed to the development of neurocomputers based on the functionalities of the brain, a seismic shift is expected in the domain of computing based on the traditional von-Neumann model. The $BrainScaleS$ \cite{schemmel2008wafer}, $SpiNNaker$ \cite{jin2010modeling} and the IBM $TrueNorth$ \cite{merolla2011digital} are instances of recent flagship neuromorphic projects that aim to develop brain-inspired computing platforms suitable for recognition (image, video, speech), classification and mining problems. While Boolean computation is based on the sequential fetch, decode and execute cycles, such neuromorphic computing architectures are massively parallel and event-driven and are potentially appealing for pattern recognition tasks and cortical brain simulations. To that end, researchers have proposed various nanoelectronic devices where the underlying device physics offer a mapping to the neuronal and synaptic operations performed in the brain. The main motivation behind the usage of such non-von Neumann post-CMOS technologies as neural and synaptic devices stems from the fact that the significant mismatch between the CMOS transistors and the underlying neuroscience mechanisms result in significant area and energy overhead for a corresponding hardware implementation. A very popular instance is the simulation of a cat's brain on IBM's Blue Gene supercomputer where the power consumption was reported to be of the order of a few $\sim MW$ \cite{adee2009ibm}. While the power required to simulate the human brain will rise significantly as we proceed along the hierarchy in the animal kingdom, actual power consumption in the mammalian brain is just a few tens of watts.

In a neuromorphic computing platform, synapses form the pathways between neurons and their strength modulate the magnitude of the signal transmitted between the neurons. The exact mechanisms that underlie the ``learning'' or ``plasticity'' of such synaptic connections are still under debate. Meanwhile, researchers have attempted to mimic several plasticity measurements observed in biological synapses in nanoelectronic devices like phase change memories \cite{jackson2013nanoscale}, $Ag-Si$ memristors \cite{jo2010nanoscale} and spintronic devices \cite{sengupta2015spin}, etc. However, majority of the research have focused on non-volatile plasticity changes of the synapse in response to the spiking patterns of the neurons it connects corresponding to long-term plasticity \cite{bi2001synaptic} and the volatility of human memory has been largely ignored. As a matter of fact, neuroscience studies performed in \cite{zucker2002short,martin2000synaptic} have demonstrated that synapses exhibit an inherent learning ability where they undergo volatile plasticity changes and ultimately undergo long-term plasticity conditionally based on the frequency of the incoming action potentials. Such volatile or meta-stable synaptic plasticity mechanisms can lead to neuromorphic architectures where the synaptic memory can adapt itself to a changing environment since sections of the memory that have been not receiving frequent stimulus can be now erased and utilized to memorize more frequent information. Hence, it is necessary to include such volatile memory transition functionalities in a neuromorphic chip in order to leverage from the computational power that such meta-stable synaptic plasticity mechanisms has to offer.

\begin{figure}[!t]
\centering
\includegraphics[width=2.9in]{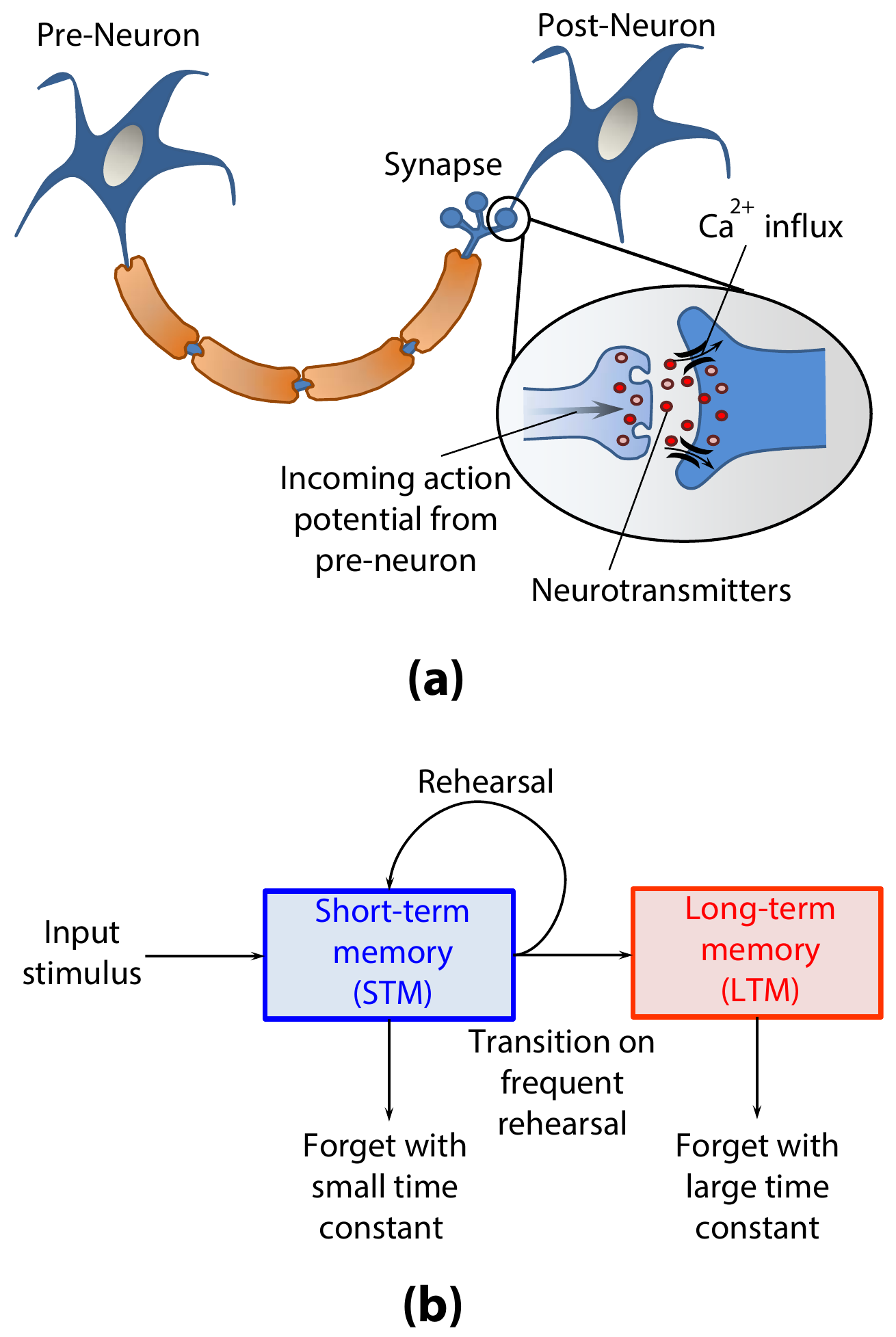}
\caption{\footnotesize{(a) A synapse is a junction joining the pre-neuron to the post-neuron. Incoming action potential from the pre-neuron results in the influx of ionic elements like $Ca^{2+}$ which, in turn, results in the release of neurotransmitters at the synaptic junction. This causes short-term synaptic plasticity (STP) while frequent action potentials result in long-term potentiation (LTP). (b) Such STP and LTP mechanisms can be related to the psychological model of human memory where memory transitions from a temporary short-term memory (STM) to a long-term memory (LTM) based on the frequency of rehearsal of the input stimulus.}}
\label{Drawing1}
\end{figure}

Fig. \ref{Drawing1} (a) demonstrates the biological process involved in such volatile synaptic plasticity changes. During the transmission of each action potential from the pre-neuron to the post-neuron through the synapse, an influx of ionic species like $Ca^{2+}, Na^{+}$ and $K^{+}$ causes the release of neurotransmitters from the pre- to the post-neuron. This results in temporary strengthening of the synaptic strength. However, in absence of the action potential, the ionic species concentration settles down to its equilibrium value and the synapse strength diminishes. This phenomenon is termed as short-term plasticity (STP) \cite{zucker2002short}. However, if the action potentials occur frequently, the concentration of the ions do not get enough time to settle down to the equilibrium concentration and this buildup of concentration eventually results in long-term strengthening of the synaptic junction. This phenomenon is termed as long-term potentiation (LTP). While STP is a meta-stable state and lasts for a very small time duration, LTP is a stable synaptic state which can last for hours, days or even years \cite{martin2000synaptic}. A similar discussion is valid for the case where there is a long-term reduction in synaptic strength with frequent stimulus and then the phenomenon is referred to as long-term depression (LTD).

Such STP and LTP mechanisms have been often correlated to the Short-Term Memory (STM) and Long-Term Memory (LTM) models proposed by Atkinson and Shiffrin \cite{atkinson1968psychology,lamprecht2004structural} (Fig. \ref{Drawing1}(b)). This psychological model partitions the human memory into an STM and an LTM. On the arrival of an input stimulus, information is first stored in the STM. However, upon frequent rehearsal, information gets transferred to the LTM. While the ``forgetting'' phenomena occurs at a fast rate in the STM, information can be stored for a much longer duration in the LTM.

In order to mimic such volatile synaptic plasticity mechanisms, a nanoelectronic device is required that is able to undergo meta-stable resistance transitions depending on the frequency of the input and also transition to a long-term stable resistance state on frequent stimulations. Hence a competition between synaptic memory reinforcement or strengthening and memory loss is a crucial requirement for such nanoelectronic synapses. In the next section, we will describe the mapping of the magnetization dynamics of a nanomagnet to such volatile synaptic plasticity mechanisms observed in the brain.
\section{Formalism}
\begin{figure*}[!t]
\centering
\includegraphics[width=6.5in]{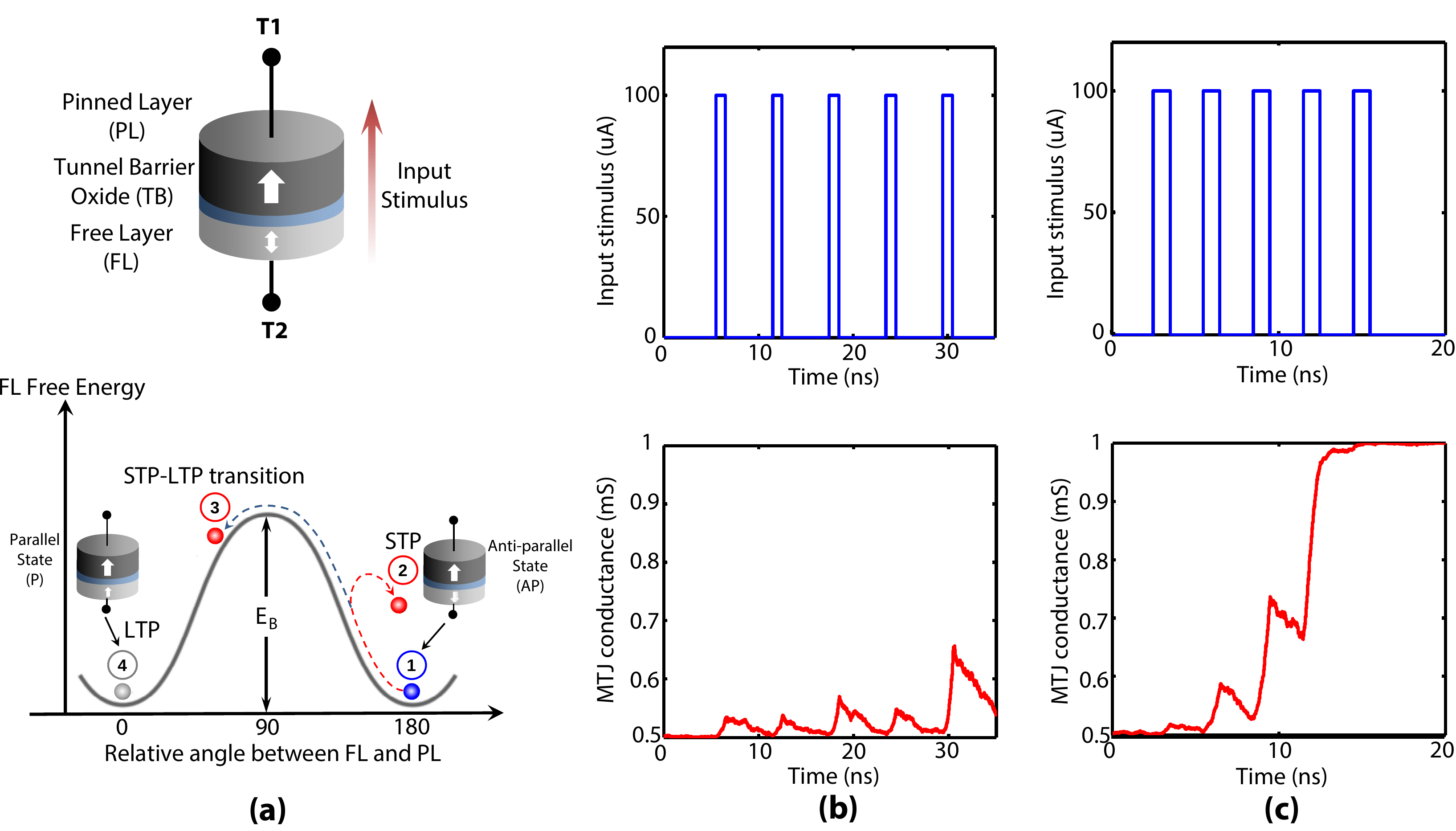}
\caption{\footnotesize{(a) An MTJ structure consists of a FL separated from a PL by a TB. Initially the MTJ synapse is in the low conductive AP state. On receiving an input stimulus it transitions to the high conductive P state conditionally depending on the time interval between the inputs. The STP-LTP behavior can be explained from the energy landscape of the FL. (b) STP behavior exhibited in the MTJ synapse. The MTJ structure was an elliptic disk of volume $\frac{\pi}{4} \times 40 \times 40 \times 1.5 nm ^3$ with saturation magnetization of $M_s = 1000 KA/m$ and damping factor, $\alpha = 0.0122$. The AP and P conductances of the MTJ were taken to be 0.5mS and 1mS respectively. The input stimulus was taken to be $100\mu A$ in magnitude (assuming $\eta=50\%$) and $1ns$ in duration. The time interval between the pulses was taken to be $6ns$. (c) The MTJ synapse undergoes LTP transition incrementally when the interval between the pulses is reduced to $3ns$.}}
\label{Drawing2}
\end{figure*}

Let us first describe the device structure and principle of operation of an MTJ \cite{julliere1975tunneling,baibich1988giant,binasch1989enhanced} as shown in Fig. \ref{Drawing2}(a). The device consists of two ferromagnetic layers separated by a tunneling oxide barrier (TB). The magnetization of one of the layers is magnetically ``pinned'' and hence it is termed as the ``pinned'' layer (PL). The magnetization of the other layer, denoted as the ``free layer'' (FL), can be manipulated by an incoming spin current $\textbf{I}_s$. The MTJ structure exhibits two extreme stable conductive states -- the low conductive ``anti-parallel'' orientation (AP), where PL and FL magnetizations are oppositely directed and the high conductive ``parallel'' orientation (P), where the magnetization of the two layers are in the same direction. 

Let us consider that the initial state of the MTJ synapse is in the low conductive AP state. Considering the input stimulus (current) to flow from terminal T2 to terminal T1, electrons will flow from terminal T1 to T2 and get spin-polarized by the PL of the MTJ. Subsequently, these spin-polarized electrons will try to orient the FL of the MTJ ``parallel'' to the PL. It is worth noting here that the spin-polarization of incoming electrons in the MTJ is analogous to the release of neurotransmitters in a biological synapse. 

The STP and LTP mechanisms exhibited in the MTJ due to the spin-polarization of the incoming electrons can be explained by the energy profile of the FL of the MTJ. Let the angle between the FL magnetization, $\widehat {\textbf {m}} $, and the PL magnetization, $\widehat {\textbf {m}}_P$, be denoted by $\theta$. The FL energy as a function of $\theta$ has been shown in Fig. \ref{Drawing2}(a) where the two energy minima points ($\theta=0^{0}$ and $\theta=180^{0}$) are separated by the energy barrier, $E_{B}$. During the transition from the AP state to the P state, the FL has to transition from $\theta=180^{0}$ to $\theta=0^{0}$. Upon the receipt of an input stimulus, the FL magnetization proceeds ``uphill'' along the energy profile (from initial point 1 to point 2 in Fig. \ref{Drawing2}(a)). However, since point 2 is a meta-stable state, it starts going ``downhill'' to point 1, once the stimulus is removed. If the input stimulus is not frequent enough, the FL will try to stabilize back to the AP state after each stimulus. However, if the stimulus is frequent, the FL will not get sufficient time to reach point 1 and ultimately will be able to overcome the energy barrier (point 3 in Fig. \ref{Drawing2}(a)). It is worth noting here, that on crossing the energy barrier at $\theta=90^{0}$, it becomes progressively difficult for the MTJ to exhibit STP and switch back to the initial AP state. This is in agreement with the psychological model of human memory where it becomes progressively difficult for the memory to ``forget'' information during transition from STM to LTM. Hence, once it has crossed the energy barrier, it starts transitioning from the STP to the LTP state (point 4 in Fig. \ref{Drawing2}(a)). The stability of the MTJ in the LTP state is dictated by the magnitude of the energy barrier. The lifetime of the LTP state is exponentially related to the energy barrier \cite{sun2005tuning}. For instance, for an energy barrier of $31.44KT$ used in this work, the LTP lifetime is $\sim 12.4$ hours while the lifetime can be extended to around $\sim 7$ years by engineering a barrier height of $40KT$. The lifetime can be varied by varying the energy barrier, or equivalently, volume of the MTJ.

The STP-LTP behavior of the MTJ can be also explained from the magnetization dynamics of the FL described by Landau-Lifshitz-Gilbert (LLG) equation with additional term to account for the spin momentum torque according to Slonczewski \cite{slonczewski1989conductance},
\begin{equation}
\label{llg}
\frac {d\widehat {\textbf {m}}} {dt} = -\gamma(\widehat {\textbf {m}} \times \textbf {H}_{eff})+ \alpha (\widehat {\textbf {m}} \times \frac {d\widehat {\textbf {m}}} {dt})+\frac{1}{qN_{s}} (\widehat {\textbf {m}} \times \textbf {I}_s \times \widehat {\textbf {m}})
\end{equation}
where, $\widehat {\textbf {m}}$ is the unit vector of FL magnetization, $\gamma= \frac {2 \mu _B \mu_0} {\hbar}$ is the gyromagnetic ratio for electron, $\alpha$ is Gilbert\textquoteright s damping ratio, $\textbf{H}_{eff}$ is the effective magnetic field including the shape anisotropy field for elliptic disks calculated using \cite{beleggia2005demagnetization}, $N_s=\frac{M_{s}V}{\mu_B}$ is the number of spins in free layer of volume $V$ ($M_{s}$ is saturation magnetization and $\mu_{B}$ is Bohr magneton), and $\textbf{I}_{s}=\eta \textbf{I}_Q$ is the spin current generated by the input stimulus $\textbf{I}_Q$ ($\eta$ is the spin-polarization efficiency of the PL). Thermal noise is included by an additional thermal field \cite{scholz2001micromagnetic}, $\textbf{H}_{thermal}=\sqrt{\frac{\alpha}{1+\alpha^{2}}\frac{2K_{B}T_{K}}{\gamma\mu_{0}M_{s}V\delta_{t}}}G_{0,1}$, where $G_{0,1}$ is a Gaussian distribution with zero mean and unit standard deviation, $K_{B}$ is Boltzmann constant, $T_{K}$ is the temperature and $\delta_{t}$ is the simulation time step. Equation ~\ref{llg} can be reformulated by simple algebraic manipulations as,
\begin{equation}
\begin{aligned}
\frac{1+\alpha^2}{\gamma} \frac {d\widehat {\textbf {m}}} {dt} =-&(\widehat {\textbf {m}} \times \textbf {H}_{eff})- \alpha (\widehat {\textbf {m}} \times \widehat {\textbf {m}} \times \textbf {H}_{eff})\\
&+\frac{1}{q\gamma N_{s}} (\alpha(\widehat {\textbf {m}} \times \textbf {I}_s )-(\widehat {\textbf {m}} \times \widehat {\textbf {m}} \times \textbf {I}_s) )
\end{aligned}
\end{equation}
Hence, in the presence of an input stimulus the magnetization of the FL starts changing due to integration of the input. However, in the absence of the input, it starts leaking back due to the first two terms in the RHS of the above equation. 

It is worth noting here that, like traditional semiconductor memories, magnitude and duration of the input stimulus will definitely have an impact on the STP-LTP transition of the synapse. However, frequency of the input is a critical factor in this scenario. Even though the total flux through the device is same, the synapse will conditionally change its state if the frequency of the input is high. We verified that this functionality is exhibited in MTJs by performing LLG simulations (including thermal noise). The conductance of the MTJ as a function of $\theta$ can be described by,
\begin{equation}
\begin{aligned}
G = G_{P}.\cos^{2}\left( \frac{\theta}{2}\right)+G_{AP}.\sin^{2}\left(\frac{\theta}{2}\right)
\end{aligned}
\end{equation}
where, $G_{P}$ ($G_{AP}$) is the MTJ conductance in the P (AP) orientation respectively. As shown in Fig. \ref{Drawing2}(b), the MTJ conductance undergoes meta-stable transitions (STP) and is not able to undergo LTP when the time interval of the input pulses is large ($6 ns$). However, on frequent stimulations with time interval as $3 ns$, the device undergoes LTP transition incrementally. Fig. \ref{Drawing2}(b) and (c) illustrates the competition between memory reinforcement and memory decay in an MTJ structure that is crucial to implement STP and LTP in the synapse.

\section{Results and Discussions}
We demonstrate simulation results to verify the STP and LTP mechanisms in an MTJ synapse depending on the time interval between stimulations. The device simulation parameters were obtained from experimental measurements \cite{pai2012spin} and have been shown in Table I.
\begin{table}[h]
\label{table}
\center
\centerline{TABLE I. Device Simulation Parameters}
\vspace{2mm}
\begin{tabular}{c c}
\hline \hline
\bfseries Parameters & \bfseries Value\\
\hline
Free layer area & $\frac{\pi}{4} \times 40 \times 40 nm^2$\\
Free layer thickness & $ 1.5 nm$\\
Saturation Magnetization, $M_{S}$ & 1000 $KA/m$ \cite{pai2012spin}\\
Gilbert Damping Factor, $\alpha$ & 0.0122 \cite{pai2012spin} \\
Energy Barrier, $E_{B}$ & 31.44 $K_{B}T$ \\
Spin polarization strength of PL, $\eta$ & 0.5 \\
MTJ conductance & 0.5-1$mS$ \\
Pulse magnitude & $100\mu A$ \\
Pulse width, $t_{PW}$ & $1ns$ \\
Temperature, $T_K$ & $300K$ \\
\hline \hline
\end{tabular}\\ 
\end{table}
\begin{figure}[!t]
\centering
\includegraphics[width=3.3in]{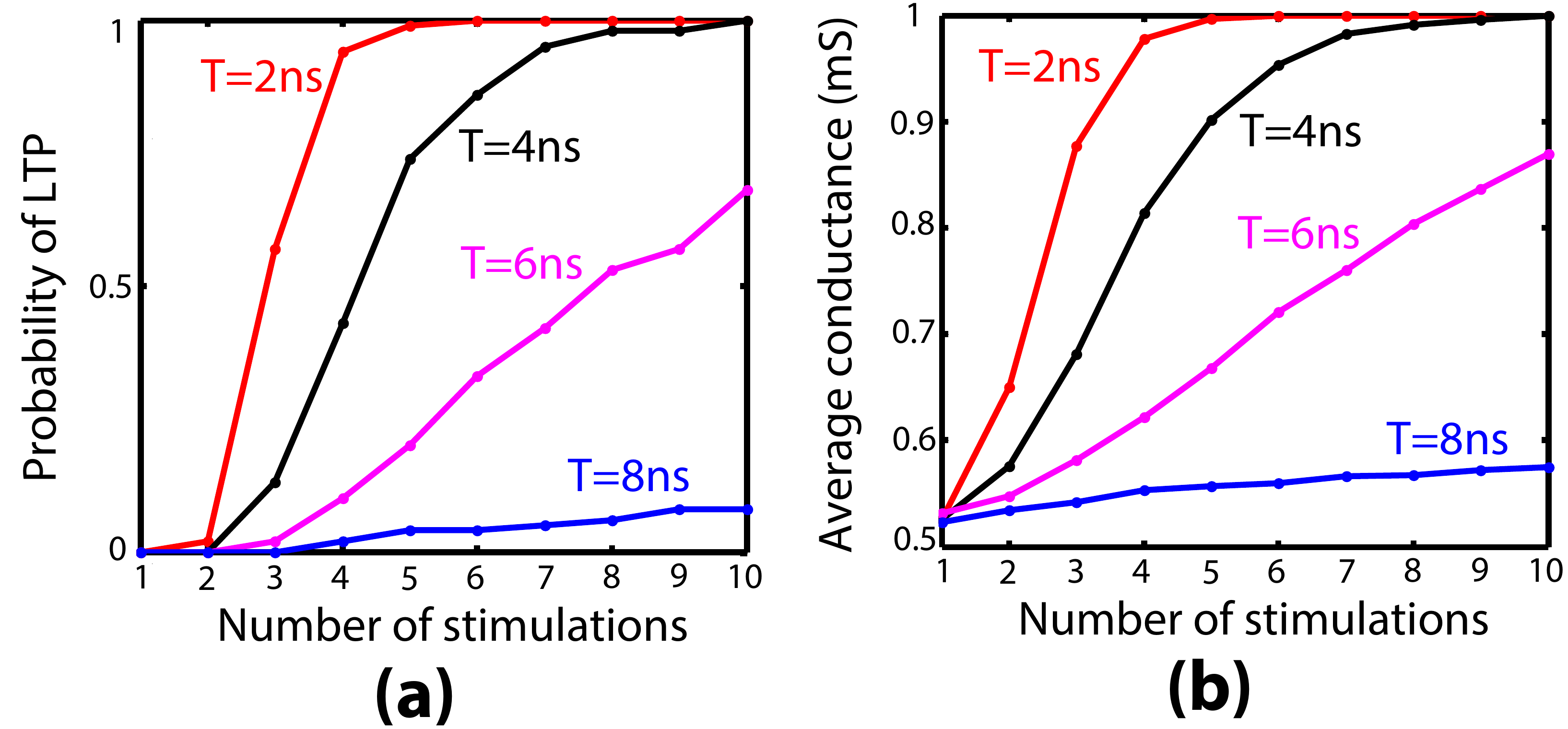}
\caption{\footnotesize{(a) Stochastic LLG simulations with thermal noise performed to illustrate the dependence of stimulation interval on the probability of LTP transition for the MTJ. The MTJ was subjected to 10 stimulations, each stimulation being a current pulse of magnitude $100 \mu A$ and $1 ns$ in duration. However, the time interval between the stimulations was varied from $2 ns$ to $8 ns$. While the probability of LTP is 1 for a time interval of $2ns$, it is very low for a time interval of $8 ns$, at the end of the 10 stimulations. (b) Average MTJ conductance plotted at the end of each stimulation. As expected, the average conductance increases faster with decrease in the stimulation interval. The results have been averaged over 100 LLG simulations.}}
\label{Drawing3}
\end{figure}
\begin{figure}[!t]
\centering
\includegraphics[width=1.9in]{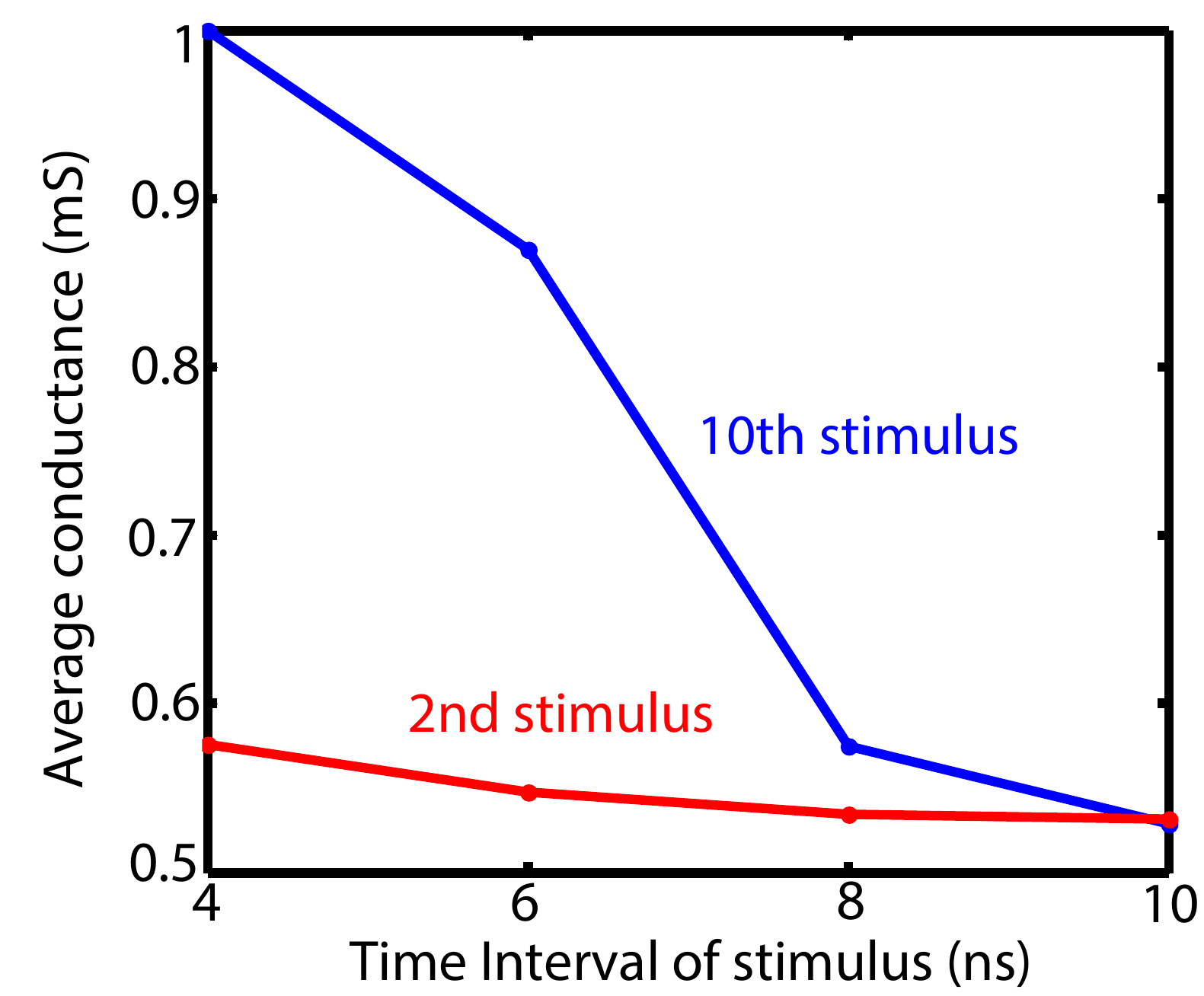}
\caption{\footnotesize{PPF (average MTJ conductance after 2nd stimulus) and PTP (average MTJ conductance after 10th stimulus) measurements in an MTJ synapse with variation in the stimulation interval. The results are in qualitative agreement to PPF and PTP measurements performed in frog neuromuscular junctions \cite{magleby1973effect,chang2011short}.}}
\label{Drawing4}
\end{figure}
\begin{figure*}[!t]
\centering
\includegraphics[width=6.6in]{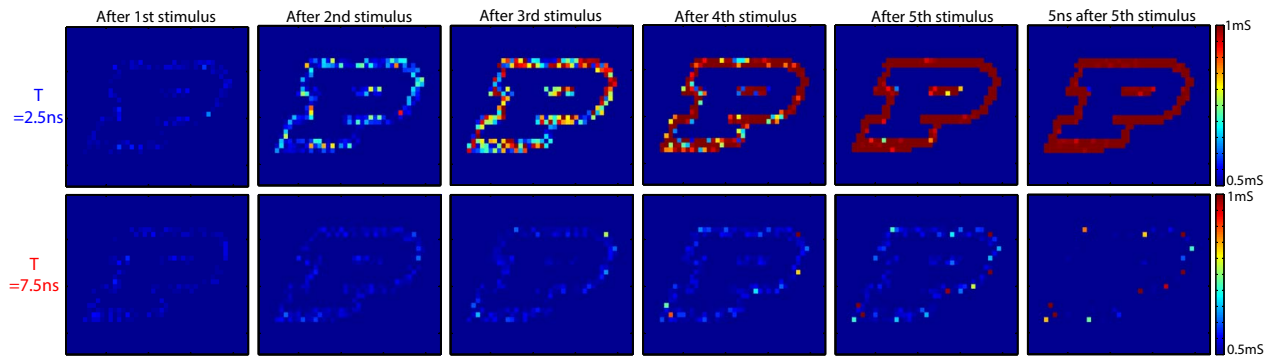}
\caption{\footnotesize{STM and LTM transition exhibited in a $34 \times 43$ MTJ memory array. The input stimulus was a binary image of the Purdue University logo where a set of 5 pulses (each of magnitude $100\mu A$ and $1ns$ in duration) was applied for each ON pixel. While the array transitioned to LTM progressively for frequent stimulations at an interval of $T=2.5ns$, it ``forgot'' the input pattern for stimulation for a time interval of $T=7.5ns$.}}
\label{Drawing5}
\end{figure*}

The MTJ was subjected to 10 stimulations, each stimulation being a current pulse of magnitude $100 \mu A$ and $1 ns$ in duration. As shown in Fig. \ref{Drawing3}, the probability of LTP transition and average device conductance at the end of each stimulation increases with decrease in the time interval between the stimulations. The dependence on stimulation time interval can be further characterized by measurements corresponding to paired-pulse facilitation (PPF: synaptic plasticity increase when a second stimulus follows a previous similar stimulus) and post-tetanic potentiation (PTP: progressive synaptic plasticity increment when a large number of such stimuli are received successively) \cite{magleby1973effect,chang2011short}. Fig. \ref{Drawing4} depicts such PPF (after 2nd stimulus) and PTP (after 10th stimulus) measurements for the MTJ synapse with variation in the stimulation interval. The measurements closely resemble measurements performed in frog neuromuscular junctions \cite{magleby1973effect} where PPF measurements revealed that there was a small synaptic conductivity increase when the stimulation rate was frequent enough while PTP measurements indicated LTP transition on frequent stimulations with a fast decay in synaptic conductivity on decrement in the stimulation rate. Hence, stimulation rate indeed plays a critical role in the MTJ synapse to determine the probability of LTP transition. 

The psychological model of STM and LTM utilizing such MTJ synapses was further explored in a $34\times 43$ memory array. The array was stimulated by a binary image of the Purdue University logo where a set of 5 pulses (each of magnitude $100\mu A$ and $1ns$ in duration) was applied for each ON pixel. The snapshots of the conductance values of the memory array after each stimulus have been shown for two different stimulation intervals of $2.5ns$ and $7.5ns$ respectively. While the memory array attempts to remember the displayed image right after stimulation, it fails to transition to LTM for the case $T=7.5ns$ and the information is eventually lost $5ns$ after stimulation. However, information gets transferred to LTM progressively for $T=2.5ns$. It is worth noting here, that the same amount of flux is transmitted through the MTJ in both cases. The simulation not only provides a visual depiction of the temporal evolution of a large array of MTJ conductances as a function of stimulus but also provides inspiration for the realization of adaptive neuromorphic systems exploiting the concepts of STM and
LTM. Readers interested in the practical implementation of such arrays of spintronic devices are referred to Ref. \cite{noguchi20157}.

\section{Conclusions}
The contributions of this work over state-of-the-art approaches may be summarized as follows. This is the first theoretical demonstration of STP and LTP mechanisms in an MTJ synapse. We demonstrated the mapping of neurotransmitter release in a biological synapse to the spin polarization of electrons in an MTJ and performed extensive simulations to illustrate the impact of stimulus frequency on the LTP probability in such an MTJ structure. There have been recent proposals of other emerging devices that can exhibit such STP-LTP mechanisms like $Ag_{2}S$ synapses \cite{ohno2011short} and $WO_{X}$ memristors \cite{chang2011short,yang2013synaptic}. However, it is worth noting here, that input stimulus magnitudes are usually in the range of volts (1.3V in \cite{chang2011short} and 80mV in \cite{ohno2011short}) and stimulus durations are of the order of a few msecs (1ms in \cite{chang2011short} and 0.5s in \cite{ohno2011short}). In contrast, similar mechanisms can be exhibited in MTJ synapses at much lower energy consumption (by stimulus magnitudes of a few hundred $\mu A$ and duration of a few $ns$). We believe that this work will stimulate proof-of-concept experiments to realize such MTJ synapses that can potentially pave the way for future ultra-low power intelligent
neuromorphic systems capable of adaptive learning.

\section*{Acknowledgements}
The work was supported in part by, Center for Spintronic Materials, Interfaces, and Novel Architectures (C-SPIN), a MARCO and DARPA sponsored StarNet center, by the Semiconductor Research Corporation, the National Science Foundation, Intel Corporation and by the National Security Science and Engineering Faculty Fellowship.

\end{document}